RESEARCH ARTICLE                                                                                            OPEN ACCESS

# Graph-Based Algorithm for a User-Aware SaaS Approach: Computing Optimal Distribution

Houda Kriouile*, Bouchra El Asri**
*(IMS Team, SIME Laboratory, ENSIAS, Mohammed V University in Rabat, Morocco)
** (IMS Team, SIME Laboratory, ENSIAS, Mohammed V University in Rabat, Morocco)

**ABSTRACT**
As a tool to exploit economies of scale, Software as a Service cloud models promote Multi-Tenancy which is the notion of sharing instances among a large group of tenants. However, Multi-Tenancy only satisfies requirements that are common to all tenants as well as the fact that tenants themselves hesitate about sharing. In a try to solve this problem, the present paper propose a User-Aware approach for Software as a Service models using Rich-Variant Components. The main contribution of this approach is a framework summarized in a graph-based algorithm enabling deduction of an optimal distribution of instances on application's tenants. To illustrate and evaluate the framework, the approach is applied on a Software as a Service Application for private school management.
***Keywords***: Algorithm, Graph Coloring, Multi-Tenancy, Rich-Variant Component, Software as a Service Applications

## I. INTRODUCTION

Cloud Computing has emerged these last decade as a new model of computing. It is nowadays one of the hottest paradigms of how to build and deliver IT services. Software as a Service (SaaS) is a form of Cloud computing that refers to software distribution model in which applications are hosted by a service provider and made availability to customers over a network. As a key enabler to exploit economies of scale, SaaS promotes Multi-Tenancy (MT), the notion of sharing resources among a large group of customer organizations, called tenants. MT brings several advantages to SaaS, however, it only satisfies requirements that are common to all tenants as well as the fact that tenants themselves hesitate about sharing.

To tackle these problems, a plethora of research work has been performed to facilitate SaaS applications customization according to the tenant-specific requirements. Most of these works are based on exploiting benefits of MT, variability management, and tenants' isolation on a single instance [1,2,3]. Likewise, our approach aims to create a flexible and reusable environment enabling greater flexibility and suppleness for customers while leveraging the economies of scale. The approach is a user-aware solution integrating a functional variability at application components level and deployment variability at multi-tenants end-users level as well. Moreover, the approach focuses on satisfying stakeholders, providers and customers, while maintaining a level of performance and remaining efficient.

The aim of our work is to provide an economy of scale for SaaS application providers while minimizing the cost to its applications tenants. We seek to achieve our goals using multi-variant components that give more possibilities of sharing allowing more instances sharing and over lower cost and better communication between tenants' communities.

This paper presents the contribution of our approach and treats the formalization of its algorithmic part. The remainder of this paper is structured as follows. Section II provides the main notion and concept making the base of knowledge of our work. Section III identifies the problem of our work as well as its motivation and its research goal. Section IV presents the main contribution of our approach consisting in a graph-based algorithm computing optimal deployment. Section V treats the algorithmic part of our approach. Section VI gives a case study illustrating our work utility. Section VII presents several approaches studied as related work. Finally, Section VIII is a conclusion of the paper.

## II. BASE OF KNOWLEDGE
**2.1. Variability-Aware System**

Variability is the capacity of a software artifact to be adapted for a specific context [4]. It can be, for example, the capacity to be extended, configured, customized, or modified. In literature, the notion of variability is largely related to Software Product Line (SPL)because it is defined in SPL context locating the differences between products of the same family. SPL community approaches focus more and more on variability resolution, and since, different definitions of variability appeared in the context of SPL. We define the variability as the description of the possible variations of a system by variation points, while a variation point identifies and locates the place where occurs the variability. It identifies possible solutions to solving this variability.

The variability can be defined at all stages of the development process. Therefore, a variability management system or software is required for all phases of system life cycle. In literature, several mechanisms are proposed for a system variability management intervening in the various phases of a system life cycle. Some examples of these mechanisms are presented below:

- Specification Phase: Iqbal, Zaidi and Murtaza propose a model for the prioritization of requirements using the Analytic Hierarchy Process [5].
- Conception Phase: Several approaches were proposed to model SPL using Feature Models, for example the Feature Oriented Domain Analysis (FODA) approach [6] that targets to capture communalities and differences at requirements level. Other approaches provide extensions to the FODA approach such as the Feature-Oriented Reuse Method (FORM) [7] whose main contribution is the decomposition of Feature Model layers to describe different perspectives.
- Testing Phase: Erwing and Walkingshaw propose the organization of the space of all variations by dimensions, which provides scoping and structuring choices [8]. They consider the "variation programming" concept for a flexible construction of all types of variation structures [8].
- Implementation Phase: Trummer proposed a corresponding data model [9] based on the Composite Application Framework (Cafe) model [1]. Applications are composed out of components that could be provided distinctly.

## 2.2. Multi-Functional Systems and Separation of concerns

The Separation of Concerns (SoC) concept was very early regarded as a key artifact to master the essential complexity of software development. It is a pragmatic application of the general strategy of "divide and rule". The underlying ideas of SoC come from E. W. Dijkstra [10]. SoC appears in the various software life cycle stages and thus it takes a variety of forms. It may be the separation in time regarding the treatment of from design to realization of the different software facets, which are then successively addressed during the development process.

Designers focus on artifacts in a reduced spectrum of concerns by using (i) generic languages (e.g. UML) or Domain Specific Languages (DSLs) sometimes called Domain Specific Modeling Languages (DSMLs) and (ii) views - targeted information encapsulation on user's business. The legitimacy of the point of views held by their intelligibility and their communicability. Indeed, an illustration of the SoC principle is the separation of "views" of a system. It can be, for example, a functional point of view describing the functional and nominal behavior of system; a fault tolerance point of view explaining the behavior in case of failure; or a performance evaluation point of view to calculate latencies, load flow, and other real-time features, of robustness models for mechanical, electromagnetic disturbance, etc. The point of view are specialized and defined with a semantic appropriate to the business domain [11].

About the architecture of a software system, more users and stakeholder, which are interested in different system aspects and its possible deployment/usage, clearly appear. Several system architectural views are defined, for example [12]. A popular approach of architectural multiviews comes from the "4+1"views methodology [13] proposed by Kruchten for the conception with UML. The point of view management irremediably brings to a consistency management issues between these views, source of many research as for example [14].

Functional domain define the main dimension of any system. They describe system activities and goals. System decomposition into a set of functional domains already existed in the field of database resulting the concept of view [15]. Multifunctional systems have been introduced to overcome problems of inconsistency and overlap between different system perspectives. The multi-functionality notion was introduced under closely related terms such as role, subject, aspect, and view, etc.

Our contribution is mainly focused on the notion of view as a mechanism of functional separation. More recently, this concept was used in service-oriented approaches to take into account the variability of service customers' needs. For example, Tran-Nguyen considers the view as a representation of a whole system from the perspective of a related set of concerns [16]. Dikanski and Abeck propose a view based approach for the specification of a service-oriented security architecture model incorporating different interrelated views in order to support the development and operation of secure service oriented applications [17].

In the context of our work, we mix the multi-functionality notion with the point of view concept as a mechanism of separation of functional concerns.

## 2.3. Cloud Computing and Multi-tenancy

The National Institute of Standards and Technology (NIST) defines the Cloud Computing as the access through a telecommunication network, by demand and self-service, to a shared pool of configurable computing resources [18]. Cloud Computing is the use of computing resources, hardware and software, which are provided as a service on a network, generally the internet. Cloud Computing loads remote services with user's data, software and computation [18].

NIST defines three main types of cloud services: Infrastructure as a Service (IaaS), Platform as a Service (PaaS), and Software as a Service (SaaS). Our work focuses on Cloud Computing SaaS services. In this type of service, applications are made available to consumers. Applications can be manipulated using a web browser. As a tool to exploit economies of scale, SaaS favors Multi-Tenancy [19].

MT is the concept of sharing resources within a large group of client organizations, called tenants. In other words, a single application instance serves multiple clients. But, although many customers use the same instance, each one has the impression that the instance is designated only for themselves. This is achieved by isolating a tenant's data from others. Unlike single-tenancy where personalization is often done by creating branches in the development tree, in MT configuration options must be integrated into the product design as in software engineering product lines. However, MT has the advantage that the infrastructure can be used as efficiently as possible to accommodate as many guests as possible on the same instance. Thus, maintenance and operating costs of the application decreases [20].

In Multi-tenant SaaS applications, variability may have different sources (evolution, maintenance, tenants requirements, etc.), but it occurs naturally [3].

### III. PROBLEM IDENTIFICATION, MOTIVATION, AND RESEARCH GOAL

#### 3.1. Problem Identification: Variability management need for Cloud environments

Cloud Computing emergence has necessitated more and more variability in the form of service types, deployment types, and the different roles of Cloud stakeholders. Thus, variability modeling is necessary to manage the complexities of cloud systems.

SaaS applications are consumed by different customers. Moreover, customers who use the same application generally have different requirements needs. Such requirements usually requires variant software architectures. In other words, when application requirements change, software architectures of these applications must be adapted to meet them. Consequently, requirements and architectures have intrinsic variability characteristics.

Furthermore, other problems are raised by MT, among other things, the need to ensure the accuracy of all possible configurations of the application. It is not enough to guarantee the accuracy of a unique configuration of an application.

On an other hand, in multi-tenant SaaS applications consumers don't have to worry about making updates and upgrades, adding security and system patches, or ensuring service availability and performance. In addition, rapid elasticity and resources pooling are essential characteristics of cloud [18], which promote the variability for cloud computing environment, especially for MT environments.

The different points cited above show the need of variability management for a cloud environment what motivated our present work benefiting from multi-functionality and MT. In this sense, our model variability will be modeled using the Multiview components as well as some graph theory concepts.

#### 3.2. Motivation by running scenario

To illustrate our model through a use case, we consider a SaaS application for a private school management accessible through a Web browser. To simplify, we reduce the application of our example into six functionalities F1 to F6 mentioned in Fig. 1. Moreover, we restrict end users of a private school management application to: administrator, professor, and student. The EGA (Education Guardianship Authority) represents the authority of education ministry and it is a special tenant that must be able to supervise schools services.

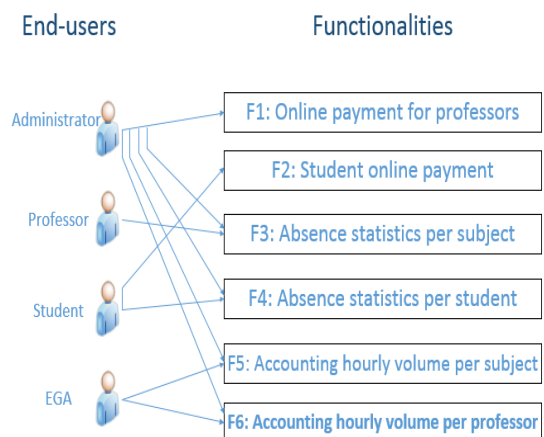

**Figure 1.** Treated application functionalities

Besides, we consider six private schools tenants of the application that are listed in Table 1. Schools which are application tenants can express their deployment requirements on sharing each application functionality.

**Table 1.** List of Schools Tenants of The Application

| School | Name | City |
|---|---|---|
| Sc1 | ABC school | Rabat |
| Sc2 | IJK school | Rabat |
| Sc3 | LMN school | Rabat |
| Sc4 | IJK school | Oujda |
| Sc5 | IJK school | Agadir |
| Sc6 | QRS school | Agadir |

### 3.3. Research Questions and Research Goal

As a key enabler to exploit economies of scale, SaaS promotes MT which brings several advantages, however, it only satisfies requirements that are common to all tenants as well as tenants themselves hesitate about sharing. So, how can we enable providers exploiting economies of scale while avoiding the problem of customers hesitation about sharing with others and allowing better communication between client communities. In the purpose of solving this problem, we need to answer the following research questions:

- Q1:How can customers' deployment requirements be captured ?
- Q2:How can deployment information be formally represented ?
- Q3:How can an optimal distribution be deduced ?

Based on the research questions, our contribution is a framework from which the information is exchanged between the provider and its customers. Our contribution, as shown in Fig. 2, can be structured into three part C1, C2, and C3 , each one dealing with one of research questions Q1, Q2, and Q3, respectively.

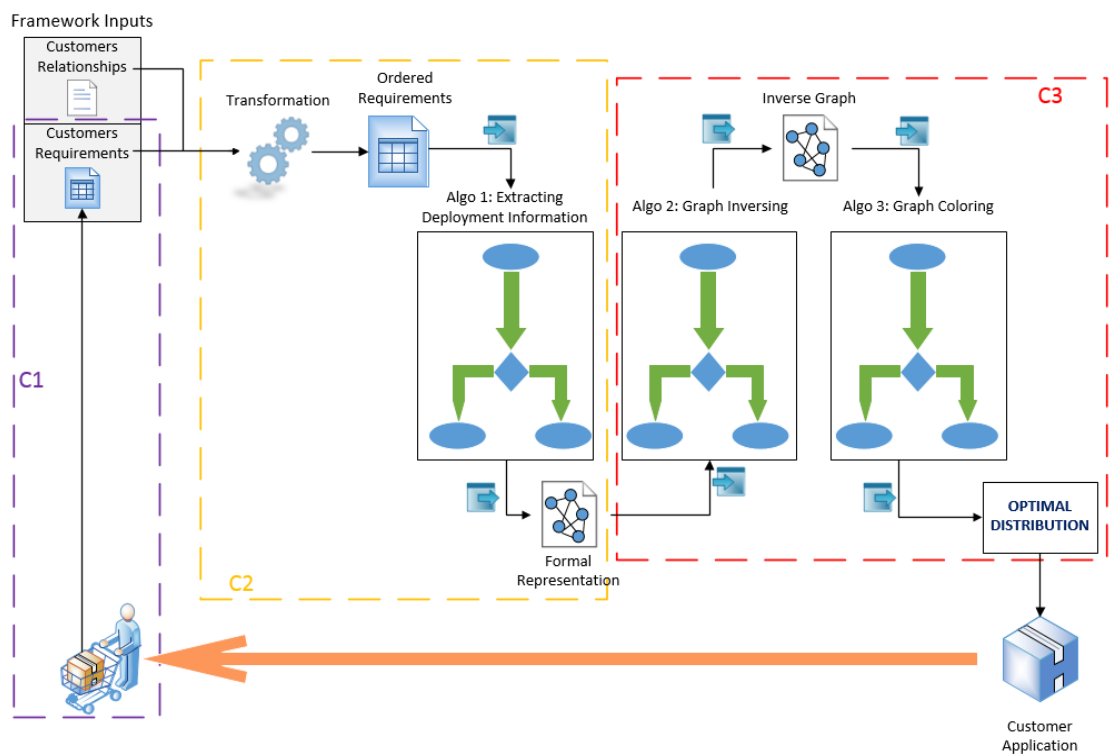

**Figure 2.** Description of our Framework

### IV. OUR CONTRIBUTION: A USER-AWARE TENANCY APPROACH BASED ON RICH-VARIANT COMPONENT

In order to provide a more flexible, more dynamic, and more reusable environment for SaaS application providers, our approach offers a users-aware tenancy based on the use of Rich-Variant Component (RVC). Through our work, we seek to exploit economies of scale while avoiding the problem of customers hesitation about sharing with others and allowing better communication between client communities.

Our approach proposes a provider platform from which the information is exchanged between the provider and its customers . The provider presents its offers and clients express their needs and requirements.

Getting by capturing tenants deployment requirements, our work aims to calculate application instances optimal distribution on tenants while respecting their deployment requirements.

In addition to client functional requirements recovery, the main idea of our work is to recover deployment-sharing requirements as well. This allows thereafter considering deployment requirements of individual tenants when calculating an application instances optimal distribution on clients of this application.

Deployment requirements expression allows tenants to express with which other tenants they wish or do not wish to share every part of the application.

A customer who pays to use an application is a tenant of this application. An application tenant may be an enterprise, a company, an association or any other organization wishing to rent the application.

Each tenant has a number of end users who are in general its employees and its staff. When designing an application, we put different roles or points of view categorizing the different users needs according to their business and missions.

In our approach, SaaS applications are built of a number of RVCs, each RVC provides atomic functionality and dynamically changes behavior according to the available user point of view. SaaS applications built based on RVC then behave differently depending on the available point of view.

The overall vision of our approach architecture is shown in Fig. 3, where all tenants use the same execution engine that executes the Rich-Variant Configurations specific to each tenant.

In the first level, the highest level of abstraction, we have the provider's catalog, which is a formal description of all available applications offered by that provider. The catalog presents applications functional variability through each application functionalities description as well as variability points specification showing thus to customers how an application can be customized. Considered as an instantiation of the catalog related with an application, the Configuration Template comes in a second level describing the RVCs that must be linked to create the specified application. Generated from a given Configuration Template, a Rich-Variant Configuration describes a specific application tailored to a specific tenant needs with a behavior that changes dynamically at runtime depending on the available end-user's role or point of view. At this level, values of the parameters or variability points of each RVC are defined, it is the description of the practical application that will be provided to the tenant.

As we have already mentioned, our SaaS applications are built from RVCs. Each RVC has a number of variants. And each application functionality is performed using a number of variants of RVCs which build the application.

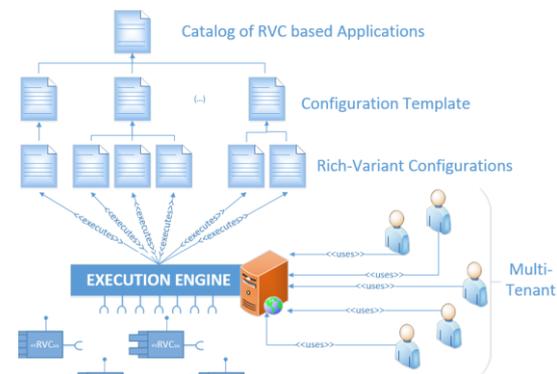

**Figure 3.** Overall architectural vision

An RVC is a Multiview component which dynamically change its behavior according to the enabled point of view. Each RVC has a number of variants that it can be deployed according of one of them each time.

From our platform, tenants choose the functionalities they desire have in the application and specify their deployment requirements for each functionality. An example of a deployment requirement is "I do not want to share the functionality F with any other tenant," or "I want to share functionality F with the tenant X" ... When a tenant doesn't precise any deployment requirement for a functionality, it means that he has no problem sharing this functionality. In this case, we consider the default value which is "Share with anyone". The next chapter shows how we formalized the expression of deployment requirements to facilitate their capture.

On customers or tenants side we talk about sharing functionalities, while on provider's side we talk about sharing RVC variants. Therefore, the initial step of our work is to translate customer requirements concerning functionalities to requirements concerning RVC variants. Two tenants can't share a functionality means that they can't share the variants involved in achieving this functionality.

Computing the optimal distribution of an application instances ends up to computing the optimal distribution of instances of RVCs building the application. The remainder of our approach is a treatment that breeds on each RVC. Thereafter, we will need deployment information of each RVC resulting from the translation of tenants requirements about functionalities and which indicate for each two tenants if they can share or not each specific RVC variant.

The representation of these deployment information is in the form of graphs, one graph for each RVC. We work with an Undirected Edge Labeled Graph. While vertices represent tenants, edges represent if two tenants can share variants or not. Besides, labels on edges indicate the variants involved in sharing relationship represented by the edge. If an edge has no label, it means that sharing relationship concerns the RVC with all its variants. Fig. 4 presents an example of deployment information represented by a graph.

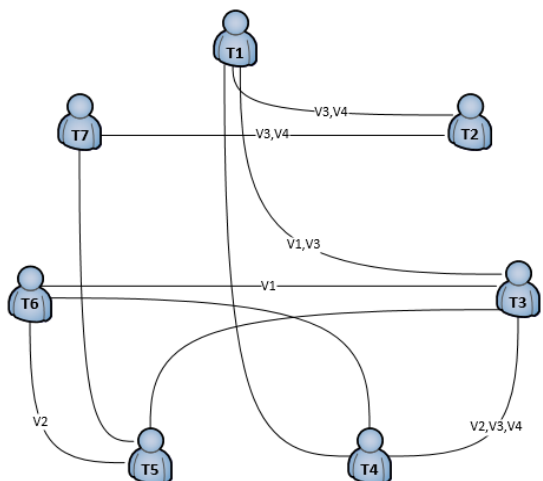

**Figure 4.** Example of deployment information graph

To derive an optimal distribution of application instances on tenants, we were inspired from well-known problems of graph theory literature [21]. Our treatment can be seen as finding a minimal clique cover of our Undirected Edge Labeled Graph. So the three steps of our treatment are as follows:

Step 1: Inverse the undirected edge labeled graph
- Keep the same vertices;
- Make each two non-adjacent vertices become adjacent with an unlabeled edge;
- Make each two adjacent and unlabeled vertices become non-adjacent;
- Make each two adjacent and labeled vertices become adjacent with a label containing the complement of variants in the initial label.

For example, for a RVC with five variants V1, V2, V3, V4, and V5, if the original label contains "V2, V5" then the label on the inverse graph is "V1, V3, V4".

Step 2: Divide vertices by RVC variants number
The second step is to divide the vertices by the number of RVC variants. If the number of variants is n, there will be n parts on each vertex each referring to a RVC variants.

Step 3: Color the Inverse Graph
The third step is to color the inverse graph. Our coloring function assigns a color to each section of each vertex so that two adjacent vertices according to a variant have different colors in the sections referring to that variant.
- Give a color to all sections of a first vertex;
- For each next vertex, for each section referring to a variant, for each color:

  o if the vertex is not adjacent to vertices of that color according to that variant, then we give it the same color;
  o if the vertex is adjacent to at least one vertex of that color, we go to next color.
- At the last color, if we didn't give any color to that section of that vertex, then we assign a new color.

This coloring part returns a set of used colors $C=\{C_1, ..., C_d\}$. Each used color is a set of sections of vertices colored by this color.

**Lemma 1:** When instantiating a RVC according to a variant, we can use the same instance according to the other variants.

Taking Lemma 1 into account, we deduce that the number of instances required to complete the deployment is the number of used colors, what means that it is the cardinality of the set $C$. Moreover, we can also deduce the optimal distribution of these instances on the different tenants, and that from the same return of the coloring function. Indeed, each color $C_k$ designates a specific instance of the RVC and the elements of this color $C_k$ refer to tenants who will use this instance and according to which variant they will use it.

In conclusion, our treatment seeking to compute valid and optimal deployment for a RVCs, can be simplified and concluded in Algorithm 0 which takes as input an Undirected Edge Labeled Graph representing deployment information about the RVC, and returns as output the set of used colors.

```
1   Algorithm 0: Overall Algorithm
2   -----------------------------------------
3   Input : G an Undirected Edge Labeled Graph,
4           and n the number of variants
5   Output :C ={C1, ..., Cd}
6   -----------------------------------------
7   1: Inverse the graph G to G'
8   2: Divide the vertices of G' by n part
9   3: Color the graph G'
10  4: return C={C1, ..., Cd}
11  -----------------------------------------
```

## V. OUR CONTRIBUTION ALGORITHMIC PART

In this chapter, we will present our work in a more formal way using formulas, algorithms and mathematical concepts.

### 5.1. Deployment requirements Capture: C1

In the aim of facilitating the capture of deployment requirements expressed by tenants, we defined four possible cases. Tenants can express their requirements for each application functionality using the following expressions:
- SWAny: Share with anyone (default value)
- SWJ(X): Share with just X ;
- DSW(X): Don't share with X ;
- DSWAny: Don't share with anyone.

Where X can take the values: "P" (as Partners), "Cp" (as Competitors), "Ti" (for a specific Tenant), or a list of the previous values.

Requirements are ordered in a table where we store the requirements of each tenant for each application functionality. We have a such table for each application. As a result of the translation of requirements concerning functionalities to requirements concerning variants, we obtain a table by RVC containing each tenant requirements for each RVC variant. However, there may be several expressions in one table cell, to settle this problem we apply the following transition rules:

- SWAny and Z give Z
- DSWAny and Z give DSWAny
- DSW(X) and DSW(Y) give DSW(X,Y)
- SWJ(X) and SWJ(Y) give DSWAny
- DSW(X) and SWJ(Y) give SWJ(Y)
- DSW(X) and SWJ(X) give DSWAny

Where Z can take any of the four possible expressions (i.e. Whatever Z).

### 5.2. From requirements to the graph: C2

From this step the work is the same for each RVC, so for the remainder of the paper we keep working on a single RVC. Then, let's have a RVC with n variants. And let m be the number of tenants. We formalize the table of m tenants Requirements about the n RVC variants by R a two dimensions (m x n) table in which each element $r_{ik}$ is the requirement of tenant i about variant k, as shown by (1):

$$R = (r_{ik}), (i=1,...,m, k=1,...,n) \quad (1)$$

The deployment information Graph is formalized by a Boolean three-dimensional matrix G (m x m x n) where the $g_{ijk}$ value indicates if tenant i and tenant j may share the variant k, as shown by (2):

$$G = (g_{ijk}), (i, j= 1,...,m, k=1,...,n) \quad (2)$$

If the $g_{ijk}$ value is 1 then both tenants i and j can share variant k, and if the $g_{ijk}$ value is 0 then they cannot share. By default, all tenants can share all variants unless they declare the opposite. Therefore, we initiate the $g_{ijk}$ values of the matrix G by 1. Thereafter, we traverse cells of requirements table R and decides whether to change the $g_{ijk}$ value according to the expression of $r_{ik}$.

- If $r_{ik}$ = DSWAny then $g_{ijk} = g_{jik} = 0$ where i and j are different.
- If $r_{ik}$ = SWJ(tenants' LIST) then $g_{ijk} = g_{jik} = 0$ where tenant j does not belong to the LIST and where i and j are different.
- If $r_{ik}$ = DSW(tenants' LIST) then $g_{ijk} = g_{jik} = 0$ where tenant j belongs to the LIST.
- If $r_{ik}$ = SWAny then we change nothing.

This step is formalized by Algorithm 1 thereafter. The end of this step makes the transition from tenant requirements to deployment information graph.

```
1   Algorithm 1 : Extracting Deployment Information
2   ------------------------------------------------
3   INPUT : Table R(m,n)
4   OUTPUT : Matrix G(m,m,n)
5   ------------------------------------------------
6       /*            Initialisation                */
7       FOR (k=0; k<n; k++)
8           FOR (i=0; i<m; i++)
9               FOR (j=0; j<m; j++)
10                  LET gijk=1
11      /*            Initialisation                */
12      FOR (k=0; k<n; k++)
13          FOR (i=0; i<m; i++)
14              IF (rik = "DSWAny") THEN
15                  FOR (j=0; j<m; j++)
16                      IF (j!=i) THEN
17                          LET gijk=0 and gjik=0
18                      ENDIF
19              ELSE IF (rik = "SWJ(LIST)") THEN
20                  FOR (j=0; j<m; j++) DO
21                      IF (j NOT IN LIST) & (j!=i) THEN
22                          LET gijk=0 and gjik=0
23                      ENDIF
24              ELSE IF (rik = "DSW(LIST)") THEN
25                  FOR (j=0; j<m; j++)
26                      IF (j IN LIST ) THEN
27                          LET gijk=0 and gjik=0
28                      ENDIF
29              ENDIF
```

### 5.3. From the graph G to its inverse: Algo.2

Thereafter, we pass from the graph G to the inverse graph formalized by a Boolean three-dimensional matrix G'(m x m x n) where the $g'_{ijk}$ value takes the opposite of $g_{ijk}$, as shown in (3):

$$G'=(g'_{ijk}), (i, j= 1,...,m, k=1,...,n) \text{ such as } g'_{ijk} = \begin{cases} 1 \text{ if } g_{ijk}= 0 \\ 0 \text{ if } g_{ijk}= 1 \end{cases} \quad (3)$$

Algorithm 2 formalize the transition from graph G to Graph G':

```
1   Algorithm 2 : Graph Inversing
2   ------------------------------
3   INPUT : Matrix G(m,m,n)
4   OUTPUT : Matrix G'(m,m,n)
5   ------------------------------
6       FOR (k=0; k<n; k++)
7           FOR (i=0; i<m; i++)
8               FOR (j=0; j<m; j++)
9                   IF (gijk=0) THEN
10                      LET g'ijk=1
11                  ELSE LET g'ijk=0
12                  ENDIF
```

### 5.4. Towards the optimal distribution: Algo.3

The optimal distribution of RVC instances is formalized by a two-dimensional matrix D (m x n) where the $d_{ik}$ value takes an integer indicating the color assigned to the part referring to the variant k from the graph vertex referring to the tenant i, as shown by (4):

$$D = (d_{ik}), (i=1,...,m, k=1,...,n) \quad (4)$$

As we had already explained in the previous chapter, to color the inverse graph we first give a first color to all parts of a first vertex. So as an initialization, we give the value 1 to all elements of the first line of the matrix D, as shown in (5):

$$d_{1k} = 1 , (k=1,...,n ) \quad (5)$$

Let h be the number of used colors, we initiate h at the value 1. And let w and u be indicators initiated to 0. Coloring of the inverse graph is completely formalized by the Algorithm 3 which takes as input the graph G' and gives as output the matrix D. The number of instances required to complete the deployment is the number of used colors, it means that it is the number h. Moreover, we can also derive the optimal distribution of these instances on the various tenants, and that from the matrix D returned by the algorithm. Indeed, each color refers to a specific instance of the RVC and the elements of the matrix D with the same value - referring to the color- show tenants who will use this instance and according to which variant they will use it.

```
Algorithm 3 : Graph Coloring
------------------------------------------------
INPUT : Matrix G'(m,m,n)
OUTPUT : Matrix D(m,n)
------------------------------------------------
    /*           Initialisation            */
    FOR (i=0; i<m; i++)
        FOR (k=0; k<n; k++)
            LET dik =1
    LET h <- 1 & w <- 0 & u <- 0
    /*           Initialisation            */
    FOR (i=1; i<m; i++)
        FOR (k=0; k<n; k++)
            LET f<-1;
            WHILE (f<h+1)
                LET j<-0;
                WHILE (j<i)
                    IF (djk = f) THEN
                        IF (g'ijk = 1) THEN
                            LET w<-1 & j<-i
                        ELSE LET j<-j+1
                        ENDIF
                    ELSE LET j<-j+1
                    ENDIF
                IF (w=0) THEN
                    LET dij<-f & u<-1 & f<-h+1
                ELSE LET f<-f+1 & w<-0
                ENDIF
            IF (u = 0) THEN
                LET h<-h+1 & dik<-h
            ELSE LET u<-0
            ENDIF
```

The following chapter includes an illustrative example to better understand and visualize the result of our approach. Moreover, in order to verify the expected results we had think about the implementation of our algorithm.

## VI. ILLUSTRATING EXAMPLE

Let us reconsider the SaaS application for a private school management initiated above. We reduced the application of our example in six functionalities F1 to F6 as mentioned in Fig. 1. In addition, we have limited the end-users in: administrator, teacher, student and EGA. The various RVCs used to make our functionalities are presented in Fig. 5. The figure illustrates the usage variants of each RVC according to the needs of end-users. The "Schedules" component has four variants A, B, C, and D, it can be used for the organization of timetables per class or per teacher, as well as for accounting hourly volume per subject or per teacher. The "Absences Monitoring" component includes two variants E and F, it can be used to account students absence or to record the current session for a teacher. The "Online Payment" component also includes two variants G and H, it can be used to make students payment or to pay part-time teachers. Finally, the "Absences Statistics" component has two variants J and K, it can be used to make absence statistics per student or per subject.

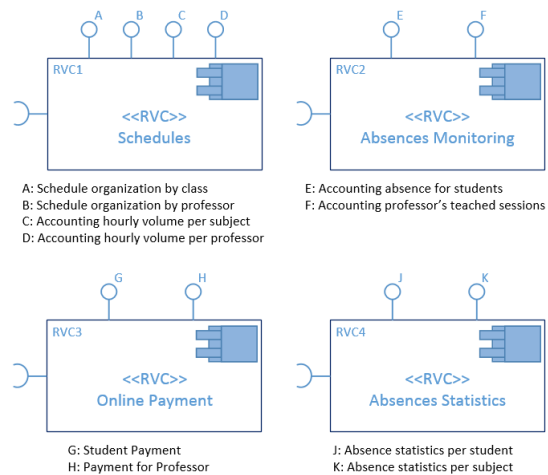

A: Schedule organization by class
B: Schedule organization by professor
C: Accounting hourly volume per subject
D: Accounting hourly volume per professor

E: Accounting absence for students
F: Accounting professor's teached sessions

G: Student Payment
H: Payment for Professor

J: Absence statistics per student
K: Absence statistics per subject

**Figure 5.** The used RVCs

Using these RVCs, we developed the Configuration Template presented in the top of Fig. 6. This template links the various RVCs needed to achieve the six functionalities of our application. Each application functionality uses a number of various RVCs variants that build the application. The figure shows the paths to achieve these functionalities as well as the users who need to perform each functionality. For example, the achievement of "F1: Online Payment For Professors" starts from the component RVC1, specifically from the second variant B of RVC1 which involves the organization of timetables by Professor and that to view timetable of teacher to pay. Then we move to the second variant F of the component RVC2 for accounting class sessions conducted by the teacher. And finally, it ends at the component RVC3, by its second variant H to make the payment of the teacher. This functionality F1 is only performed by an administrator.

As shown in Fig. 6, the functionality "F3: Absence statistics per subject " is performed by the teacher in order to assess the presence in its own subjects, as it is performed by an administrator to monitor the progress of the various school subjects. Similarly, the functionality "F4: Absence statistics per student" is performed by the administrator and the student each for its own purpose. The functionality "F2: Student Online Payment" is done exclusively by the student. Both functionalities "F5:

Accounting hourly volume per subject" and "F6: Accounting hourly volume per professor" are performed by the administrator or by the EGA users to control school services. Both of these functionalities are realized by the third (C) and the fourth (D) variants of RVC1.

In general, a school does not wish to share with its competitors that it may specify or can be defined as the schools of same type (primary school, middle school, high school, vocational training school, college ...) from the same town.

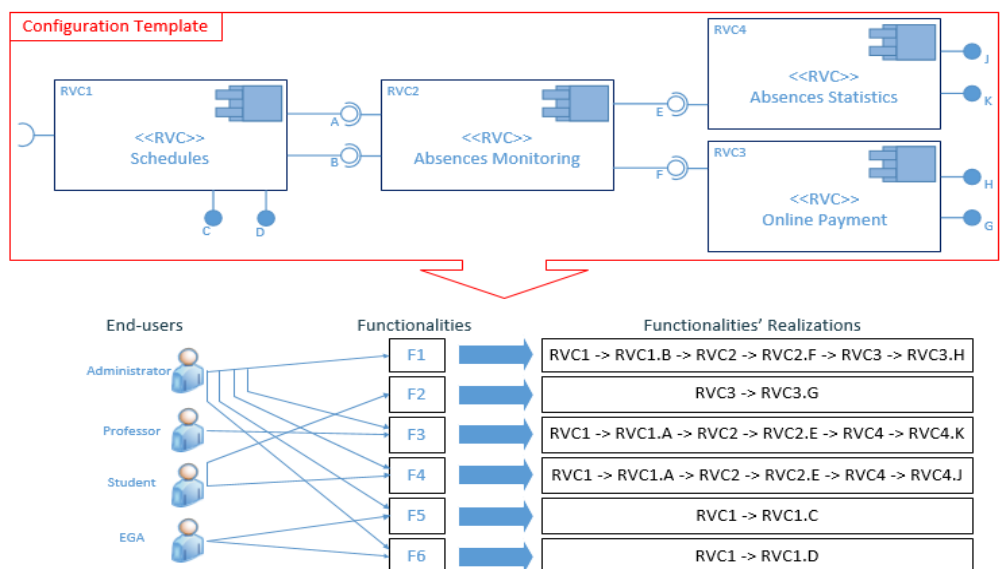

**Figure 6.** Configuration Template achieving functionalities

**Table 2.** Deployment Requirements Expressed By The Six Tenants Concerning The Six Application Functionalities

| Feature | Variant | Sc1 | Sc2 | Sc3 | Sc4 | Sc5 | Sc6 |
|---|---|---|---|---|---|---|---|
| F1 | B, F, H | DSWAny | DSW(Sc3) | DSW(Cp, Sc6) | DSW(P) | DSW(Sc3) | SWAny |
| F2 | G | DSWAny | SWJ(P) | ---------- | ---------- | ---------- | SWAny |
| F3 | A, E, K | DSWAny | ---------- | DSW(Cp) | SWJ(P) | ---------- | SWAny |
| F4 | A, E, J | DSWAny | ---------- | DSW(Sc4) | SWJ(P) | DSW(Sc2) | SWAny |
| F5 | C | DSWAny | ---------- | ---------- | ---------- | ---------- | SWAny |
| F6 | D | DSWAny | DSW(Cp) | DSW(Sc6) | SWJ(P) | DSW(Cp) | SWAny |

Also a school may wish to share instances with its partners to collaborate in their work. The partners of a school are, in general, schools of the same group of schools located in other cities, in addition to schools in partnership mentioned by the school tenant of the application. The schools of the same group may, for example, wish to share the instance of the component "Absences Statistics" to compare and analyze the results. On the other hand, schools have to share instances of the component "Schedules" with the EGA to enable it to monitor schools through both F5 and F6 functionalities accounting the hourly volumes. The application used by the EGA may be different from those used by schools (less functionalities), but it must at least contain the component "Schedules".

Application tenants schools express their deployment requirements on sharing a specific application functionality. Tenants expression of deployment requirements concerning application functionalities is technically translated in deployment requirements concerning variants of application RVCs.

According to Competitors and Partners definitions mentioned previously, the relationships between the six private schools tenants of the application listed in Table 1 are: Sc1, Sc2, and Sc3 are competitors; Sc2, Sc4, and Sc5 are partners; Sc5 and Sc6 are competitors. Tenants deployment requirements concerning the illustrating example are presented in Table 2. Each tenant expresses its requirements for each functionality, otherwise it means that the tenant has no problems to share with other tenants. Thus, the empty cells of the table take the default value, which is SWAny.

The initial step is to translate requirements about functionalities to requirements about RVCs variants. Using the transition rules cited in the previous chapter and detailing lists of tenants partners and competitors, we pass from Table 2 to Table 3 which includes four tables each for a RVC.

Table 3. Deployment Requirements Concerning Application Rvcs Variants

| RVC | Variant | Sc1 | Sc2 | Sc3 | Sc4 | Sc5 | Sc6 |
|---|---|---|---|---|---|---|---|
| 1 | A | DSWAny | SWAny | DSW(T1,T2,T4) | DSW(T2,T5) | DSW(T2) | SWAny |
| | B | DSWAny | DSW(T3) | DSW(T1,T2,T6) | DSW(T2,T5) | DSW(T3) | SWAny |
| | C | DSWAny | SWAny | SWAny | SWAny | SWAny | SWAny |
| | D | DSWAny | DSW(T1,T3) | DSW(T6) | SWJ(T2,T5) | DSW(T6) | SWAny |
| 2 | E | DSWAny | SWAny | DSW(T1,T2,T4) | DSW(T2,T5) | DSW(T2) | SWAny |
| | F | DSWAny | DSW(T3) | DSW(T1,T2,T6) | DSW(T2,T5) | DSW(T3) | SWAny |
| 3 | G | DSWAny | SWJ(P) | SWAny | SWAny | SWAny | SWAny |
| | H | DSWAny | DSW(T3) | DSW(T1,T2,T6) | DSW(T2,T5) | DSW(T3) | SWAny |
| 4 | J | DSWAny | SWAny | DSW(T4) | SWJ(T2,T5) | DSW(T2) | SWAny |
| | K | DSWAny | SWAny | DSW(T1,T2) | SWJ(T2,T5) | SWAny | SWAny |

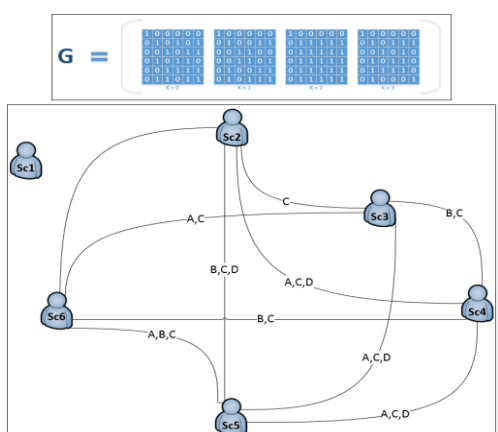

**Figure 7.** Deployment information graph concerning the RVC1 resulting from the use of our algorithm

To simplify the illustration of our algorithms, we focus on a single RVC - the same work is done for the other RVCs - and we will just give the results for the other RVCs. So, for the illustration of the different remaining steps of the algorithm, we consider the first component of Fig. 5, the RVC1 named "Schedules". This component has four variants. The framed portion of Table 3 shows requirements concerning variants of RVC1. We take this portion as input of our algorithm, it is the Table R. The algorithm deduces the matrix G. Fig. 7 shows the numerical values of G elements as well as its graphical representation.

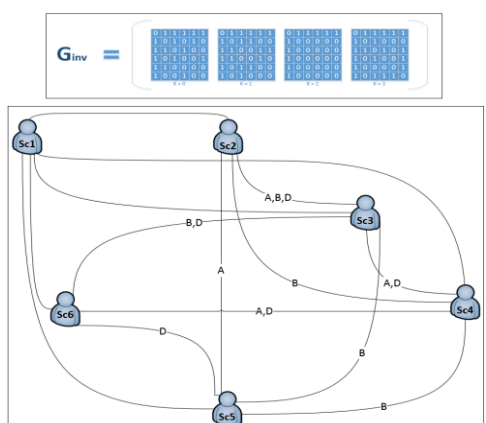

**Figure 8.** Inverse graph of deployment information graph concerning the RVC1

The next step is to inverse the graph G to obtain the graph G '. The resulting inverse graph is shown in Fig. 8 in the form of a numerical matrix and in the form of an Undirected Edge Labeled Graph.

The final step is to apply Algorithm 3 to color the inverse graph. The algorithm takes as input the matrix G' presented in Fig.8 and gives as output the matrix D de dimension (6 x 4). The result obtained by the application of the Algorithm 3is presented in Fig. 9. We have the information for each tenant which RVC instance should get according to each variant.

|   | A | B | C | D |
|---|---|---|---|---|
| Sc1 | 1 | 1 | 1 | 1 |
| Sc2 | 2 | 2 | 2 | 2 |
| Sc3 | 3 | 3 | 2 | 3 |
| Sc4 | 2 | 3 | 2 | 2 |
| Sc5 | 3 | 2 | 2 | 2 |
| Sc6 | 3 | 2 | 2 | 4 |

**Figure 9.** Output of Algorithm 3 application

From Algorithm 3 output we deduce the optimal distribution of RVC1 instances exposed in Table 4. Each number from Fig. 9 refers to an instance, for example, instance number 1of RVC1 must be given to tenant Sc1 only and according to all variant.

**Table 4.** RVC1 instances distribution resulting from the algorithm

| \Instance Variant | I1 | I2 | I3 | I4 |
|---|---|---|---|---|
| A | Sc1 | Sc2, Sc4 | Sc3, Sc5, Sc6 | ---- |
| B | Sc1 | Sc2, Sc5, Sc6 | Sc3, Sc4 | ---- |
| C | Sc1 | Sc2, Sc3, Sc4, Sc5, Sc6 | ---- | ---- |
| D | Sc1 | Sc2, Sc4, Sc5 | Sc3 | Sc6 |

As regards the other RVCs, instances distribution resulting from the application of our algorithms is presented in Table 5. Only three instances are needed for RVC2 and RVC4. And a more instance is necessary for the RVC3 but only according to variant H. So, for the six tenants, we only need four instances to respect all tenants requirements about deployment and sharing functionalities.

**Table 5.** RVCs instances distribution resulting from algorithms application

| RVC | Variant\ Instance | I1 | I2 | I3 | I4 |
|---|---|---|---|---|---|
| RVC1 | A | T1 | T2, T4 | T3, T5, T6 | ---- |
|  | B | T1 | T2, T5, T6 | T3, T4 | ---- |
|  | C | T1 | T2, T3, T4, T5, T6 | ---- | ---- |
|  | D | T1 | T2, T4, T5 | T3 | T6 |
| RVC2 | E | T1 | T2, T4 | T3, T5, T6 | ---- |
|  | F | T1 | T2, T5, T6 | T3, T4 | ---- |
| RVC3 | G | T1 | T2, T5 | T3, T4, T6 | ---- |
|  | H | T1 | T2, T4, T6 | T3 | T5 |
| RVC4 | J | T1 | T2, T3, T6 | T4, T5 | ---- |
|  | K | T1 | T2, T4, T5 | T3, T6 | ---- |

## VII. RELATED WORK

Several works have been performed to address the realization and variability of Multi-tenancy systems in general and Multi-tenancy SaaS applications in particular. In [22], the authors propose a SaaS customization policy as well as a supporting framework that is realized through a design-time tooling and a run-time environment. However, this work mainly focuses on the unique issues in service customization for a given set of requirements. Reference [23] is an example of several works that addresses the challenge of introducing flexibility into Multi-Tenancy applications. Its authors discus the configuration issues and challenges related to it, and propose a competency model and a methodology framework that both aim to support SaaS providers in planning and evaluating their configuration and customizing strategies. In [24], the authors use a directed hypergraph based service model to represent hierarchical services and Multi-Tenancy applications. Based on these graphs, it is possible to represent dependencies between services and application structures from which Multi-Tenancy applications can be constructed fulfilling customer requirements.

Several research works have been performed in the context of architectural patterns for developing and deploying customizable multi-tenant applications for Cloud environment. Several approaches from those - cited below - was studied and compared in Table 6. The comparison is based on common characteristics shared by the studied approaches.

**Approach A:** (Composite as a Service (CaaS) [1][25]) show how applications built of components, using different Cloud service models, can be composed to form new applications that can be offered as a new service.

**Approach B:** (Matchmaking of IaaS Offers Leveraging Linked Data [2][26]) present models of Expressive Search Requests and Service Offer Descriptions allowing matchmaking of highly configurable services that are dynamic and depend on request.

**Approach C:** (Service line engineering [3]) present an integrated service engineering method, that supports co-existing tenant-specific configurations and that facilitates the development and management of customizable, multi-tenant SaaS applications.

**Approach D:** (Mixed-tenancy Systems [19]) addresses the deployment variability based on the SaaS tenants requirements about sharing infrastructure, application codes or data with other tenants. It proposes a hybrid solution between multi-tenancy and simple tenancy.

The new notion brought by our approach and that is not proposed by the others approaches is the roles accessibility based on the concept of Multiview. All cited approaches aim to improve flexibility and reusability in their ways. To exploit economies of scale some approaches rely on the multi-tenancy, we do the same in our approach but in addition we benefit from the use of Multiview notion to exploit more and more economies of scale.

Table 6. A Comparative Study On Customizable Approaches For Cloud Environment

| Approaches | Composite as a Service Approach | Matchmaking of IaaS Offers Leveraging Linked Data Approach | Service line engineering Approach | Mixed-tenancy Systems Approach | Our Approach |
|---|---|---|---|---|---|
| Cloud application area | SaaS | IaaS, Service Computing | SaaS | SaaS | SaaS |
| Variability | -Functional -Deployment | Deployment | Functional | -Deployment -Functional | -Functional -Deployment |
| Accessibility by roles | Not proposed | Not proposed | Not proposed | Not proposed | Use of Multiview concept |
| Flexibility | Dynamically scale based on customer demand | Service consumer might specify a flexible search request using enumerations and ranges | Use of Service line and Workflows | Flexibility to use depending to the tenant using the application | Flexibility according to tenants, and flexibility according to enabled view |
| Reusability | Use of component-based software | Service Variant Hierarchy promotes reuse | Modular middleware layer | Use of application component | Use of RVCs |
| Economies of scale | Use of highly flexible templates enabling increasing customers base | Not proposed | Application-level multi-tenancy | Mixed tenancy (hybrid solution between multi-tenancy and simple tenancy) | - Multi-tenancy - Multiview notion |

## VIII. CONCLUSION

Flexibility and reusability are challenging issues for multi-tenancy SaaS applications. In this regard, our user-aware SaaS approach consists in integrating two types of variability to create a more flexible and reusable SaaS environment while exploiting economies of scale and avoiding the problem of tenants hesitation about sharing with others. In this context, this paper addresses the algorithmic part formalization, which aims to compute a valid and optimal RVC instances distribution on tenants while respecting their deployment requirements. For this purpose, we first presented the context and motivations of the problem. Then, we presented our User-Aware SaaS Approach. Then, we treated the formalization of our approach using some mathematics concepts. Finally, to illustrate our model we applied our algorithm to a case study. As future work, we think about projecting our approach in the domain of Model-driven engineering for a more modern and more general vision.